\documentclass[a4paper]{jpconf}
\usepackage{graphicx}
\usepackage{color}
\definecolor{GreenYellow}   {cmyk}{0.15,0,0.69,0}
\definecolor{Yellow}        {cmyk}{0,0,1,0}
\definecolor{Goldenrod}     {cmyk}{0,0.10,0.84,0}
\definecolor{Dandelion}     {cmyk}{0,0.29,0.84,0}
\definecolor{Apricot}       {cmyk}{0,0.32,0.52,0}
\definecolor{Peach}         {cmyk}{0,0.50,0.70,0}
\definecolor{Melon}         {cmyk}{0,0.46,0.50,0}
\definecolor{YellowOrange}  {cmyk}{0,0.42,1,0}
\definecolor{Orange}        {cmyk}{0,0.61,0.87,0}
\definecolor{BurntOrange}   {cmyk}{0,0.51,1,0}
\definecolor{Bittersweet}   {cmyk}{0,0.75,1,0.24}
\definecolor{RedOrange}     {cmyk}{0,0.77,0.87,0}
\definecolor{Mahogany}      {cmyk}{0,0.85,0.87,0.35}
\definecolor{Maroon}        {cmyk}{0,0.87,0.68,0.32}
\definecolor{BrickRed}      {cmyk}{0,0.89,0.94,0.28}
\definecolor{Red}           {cmyk}{0,1,1,0}
\definecolor{OrangeRed}     {cmyk}{0,1,0.50,0}
\definecolor{RubineRed}     {cmyk}{0,1,0.13,0}
\definecolor{WildStrawberry}{cmyk}{0,0.96,0.39,0}
\definecolor{Salmon}        {cmyk}{0,0.53,0.38,0}
\definecolor{CarnationPink} {cmyk}{0,0.63,0,0}
\definecolor{Magenta}       {cmyk}{0,1,0,0}
\definecolor{VioletRed}     {cmyk}{0,0.81,0,0}
\definecolor{Rhodamine}     {cmyk}{0,0.82,0,0}
\definecolor{Mulberry}      {cmyk}{0.34,0.90,0,0.02}
\definecolor{RedViolet}     {cmyk}{0.07,0.90,0,0.34}
\definecolor{Fuchsia}       {cmyk}{0.47,0.91,0,0.08}
\definecolor{Lavender}      {cmyk}{0,0.48,0,0}
\definecolor{Thistle}       {cmyk}{0.12,0.59,0,0}
\definecolor{Orchid}        {cmyk}{0.32,0.64,0,0}
\definecolor{DarkOrchid}    {cmyk}{0.40,0.80,0.20,0}
\definecolor{Purple}        {cmyk}{0.45,0.86,0,0}
\definecolor{Plum}          {cmyk}{0.50,1,0,0}
\definecolor{Violet}        {cmyk}{0.79,0.88,0,0}
\definecolor{RoyalPurple}   {cmyk}{0.75,0.90,0,0}
\definecolor{BlueViolet}    {cmyk}{0.86,0.91,0,0.04}
\definecolor{Periwinkle}    {cmyk}{0.57,0.55,0,0}
\definecolor{CadetBlue}     {cmyk}{0.62,0.57,0.23,0}
\definecolor{CornflowerBlue}{cmyk}{0.65,0.13,0,0}
\definecolor{MidnightBlue}  {cmyk}{0.98,0.13,0,0.43}
\definecolor{NavyBlue}      {cmyk}{0.94,0.54,0,0}
\definecolor{RoyalBlue}     {cmyk}{1,0.50,0,0}
\definecolor{Blue}          {cmyk}{1,1,0,0}
\definecolor{Cerulean}      {cmyk}{0.94,0.11,0,0}
\definecolor{Cyan}          {cmyk}{1,0,0,0}
\definecolor{ProcessBlue}   {cmyk}{0.96,0,0,0}
\definecolor{SkyBlue}       {cmyk}{0.62,0,0.12,0}
\definecolor{Turquoise}     {cmyk}{0.85,0,0.20,0}
\definecolor{TealBlue}      {cmyk}{0.86,0,0.34,0.02}
\definecolor{Aquamarine}    {cmyk}{0.82,0,0.30,0}
\definecolor{BlueGreen}     {cmyk}{0.85,0,0.33,0}
\definecolor{Emerald}       {cmyk}{1,0,0.50,0}
\definecolor{JungleGreen}   {cmyk}{0.99,0,0.52,0}
\definecolor{SeaGreen}      {cmyk}{0.69,0,0.50,0}
\definecolor{Green}         {cmyk}{1,0,1,0}
\definecolor{ForestGreen}   {cmyk}{0.91,0,0.88,0.12}
\definecolor{PineGreen}     {cmyk}{0.92,0,0.59,0.25}
\definecolor{LimeGreen}     {cmyk}{0.50,0,1,0}
\definecolor{YellowGreen}   {cmyk}{0.44,0,0.74,0}
\definecolor{SpringGreen}   {cmyk}{0.26,0,0.76,0}
\definecolor{OliveGreen}    {cmyk}{0.64,0,0.95,0.40}
\definecolor{RawSienna}     {cmyk}{0,0.72,1,0.45}
\definecolor{Sepia}         {cmyk}{0,0.83,1,0.70}
\definecolor{Brown}         {cmyk}{0,0.81,1,0.60}
\definecolor{Tan}           {cmyk}{0.14,0.42,0.56,0}
\definecolor{Gray}          {cmyk}{0,0,0,0.50}
\definecolor{Black}         {cmyk}{0,0,0,1}
\definecolor{White}         {cmyk}{0,0,0,0}

\usepackage{colordvi}
\usepackage{axodraw}
\begin{document}

\begin{flushright}
MAN/HEP/2009/14\\
April 2009
\end{flushright}

\title{The Little Review on Leptogenesis}

\author{Apostolos Pilaftsis}

\address{School of Physics and Astronomy, University of Manchester,
Manchester M13 9PL, United Kingdom}

\ead{apostolos.pilaftsis@manchester.ac.uk}

\begin{abstract} This is a brief review on the scenario of baryogenesis
through leptogenesis.  Leptogenesis is  an appealing scenario that may
relate the observed baryon asymmetry in the Universe to the low-energy
neutrino data.   In this  review talk, particular  emphasis is  put on
recent  developments on  the  field, such  as  the flavourdynamics  of
leptogenesis  and  resonant leptogenesis  near  the electroweak  phase
transition. It is illustrated how these recent developments enable the
modelling of phenomenologically predictive scenarios that can directly
be  tested at  the LHC  and  indirectly in  low-energy experiments  of
lepton-number and lepton-flavour violation.
\end{abstract}

\section{Introduction}
\medskip

An  elegant  framework to  consistently  address  the observed  Baryon
Asymmetry     in     the     Universe     (BAU)~\cite{ADS,WMAP}     is
leptogenesis~\cite{FY}.   According   to  the  standard   paradigm  of
leptogenesis~\cite{review},   minimal  extensions   of   the  Standard
Model~(SM) realize  heavy Majorana neutrinos of masses  related to the
Grand Unified Theory (GUT) scale $M_{\rm GUT} \sim 10^{16}$ that decay
out  of  equilibrium  and  create   a  net  excess  of  lepton  number
$(L)$.  This excess  in $L$  gets then  reprocessed into  the observed
baryon   number  $(B)$,   through   the  $(B+L)$-violating   sphaleron
interactions~\cite{KRS}.  The attractive feature of such a scenario is
that  the GUT-scale heavy  Majorana neutrinos  could also  explain the
observed smallness in  mass of the SM light neutrinos  by means of the
so-called seesaw mechanism~\cite{seesaw}.

The  original  GUT-scale  leptogenesis  scenario, however,  runs  into
certain difficulties, when one attempts to explain the flatness of the
Universe and  other cosmological data~\cite{WMAP}  within supergravity
models   of  inflation.    To  avoid   overproduction   of  gravitinos
$\widetilde{G}$ whose late decays  may ruin the successful predictions
of  Big Bang  Nucleosynthesis  (BBN), the  reheat temperature  $T_{\rm
reh}$  of the Universe  should be  lower than  $10^9$--$10^6$~GeV, for
$m_{\widetilde{G}} =  8$--0.2~TeV~\cite{gravitino}.  This implies that
the heavy Majorana neutrinos should  accordingly have masses as low as
$T_{\rm  reh} \stackrel{<}{{}_\sim}  10^9$~GeV, thereby  rendering the
relation of  these particles with GUT-scale physics  less natural.  On
the  other  hand, it  proves  very  difficult  to directly  probe  the
heavy-neutrino  sector  of  such  a model  at  high-energy  colliders,
e.g.~at the LHC or ILC, or in any other foreseeable experiment.

A  potentially  interesting solution  to  the  above  problems may  be
obtained    within   the    framework    of   resonant    leptogenesis
(RL)~\cite{APRD}.  The  key aspect of  RL is that  self-energy effects
dominate  the  leptonic  asymmetries~\cite{LiuSegre}, when  two  heavy
Majorana neutrinos happen to have a small mass difference with respect
to their actual masses.  If this mass difference becomes comparable to
the  heavy neutrino  widths, a  resonant enhancement  of  the leptonic
asymmetries    takes   place    that   may    reach    values   ${\cal
O}(1)$~\cite{APRD,PU}.  An indispensable feature  of RL models is that
flavour   effects   due   to   the  light-to-heavy   neutrino   Yukawa
couplings~\cite{APtau,PU2,EMX} play a dramatic role and can modify the
predictions for the BAU  by many orders of magnitude~\cite{APtau,PU2}.
Most    importantly,     these    flavour    effects     enable    the
modelling~\cite{PU2}  of minimal  RL scenarios  with electroweak-scale
heavy   Majorana    neutrinos   that   could   be    tested   at   the
LHC~\cite{APZPC,DGP,Kersten/Smirnov}  and   in  other  non-accelerator
experiments, while maintaining  agreement with the low-energy neutrino
data.    Many   variants   of   RL   have   been   proposed   in   the
literature~\cite{RLpapers,RLextra},           including           soft
leptogenesis~\cite{soft} and radiative leptogenesis~\cite{rad}.

In this review  we give a brief exposition  of selected topics related
to   recent    advancements   of    the   field.    In    detail,   in
Section~\ref{sec:matter}, we  remind ourselves about  Sakharov's basic
conditions for  baryogenesis and list  a few classical  scenarios that
have  been suggested  in the  literature, including  leptogenesis.  In
Section~\ref{sec:flavour}  we discuss  the importance  of  the various
sources  of  flavour effects  on  leptogenesis.  Section  \ref{sec:RL}
presents  the  field theory  of  RL,  including  a discussion  of  its
flavourdynamics.    Section~\ref{sec:PPP}   discusses  the   potential
particle-physics implications of RL.  Finally, Section~\ref{sec:Concl}
summarizes the main points of this review.

\section{Matter--Antimatter Asymmetry and Leptogenesis \label{sec:matter}}
\medskip

There  are two  pieces of  observational evidence  that  establish the
domination   of  the   observable  matter   over  antimatter   in  our
Universe. The first  stems from the analysis of  the power spectrum of
the  cosmic  microwave   background~\cite{WMAP}.  In  particular,  the
baryon-to-photon ratio $\eta_B$ of number densities is found to be
\begin{equation}
  \label{etaCMB}
\eta^{\rm CMB}_B \ =\ \frac{n_B}{n_\gamma}\ =\ 
6.1^{+0.3}_{-0.2}\,\times\,10^{-10}\; .
\end{equation}
The second piece of evidence originates from the concordance of the
light elements and big bang nucleosynthesis, which gives~\cite{PDG}
\begin{equation}
  \label{etaBBN}
\eta^{\rm BBN}_B\ =\ 
3.4\mbox{--}6.9\,\times\,10^{-10}\; .
\end{equation}
In  spite  of their  different  origin,  the  two deduced  values  for
$\eta_B$ are in good agreement with one another. 

More  than  fifty  years  ago~\cite{ADS},  Sakharov  has  studied  the
conditions  for   which  a   baryon  asymmetry  may   successfully  be
generated. These basic conditions are:
\begin{itemize}

\item the theory should have B-violating interactions.

\item the interactions should violate both C and CP.

\item the processes of a net baryon-number generation should have a
 degree of irreversibility due to an out-of-thermal equilibrium
 dynamics.

\end{itemize}

There  are many  scenarios that  have  been proposed  in the  existing
literature. A few classical examples are:

\begin{itemize}

\item 
{\it  Baryogenesis  through the  decay  of  a  heavy particle.}   This
scenario is  based on the out-of-equilibrium $B$-violating  decay of a
heavy GUT  particle, e.g.  in SO(10)  theory~\cite{GUTbau}.  A serious
constraint  difficult to overcome  in these  scenarios comes  from the
strict experimental lower limit on proton's long lifetime~\cite{PDG}.

\item 
{\it  Baryogenesis  at the  electroweak  phase  transition}.  In  this
scenario, BAU is generated by $(B+L)$-violating sphaleron interactions
at  $T\sim   T_c  \approx  140$~GeV,  through  a   first  order  phase
transition~\cite{KRS}.   However, such a  scenario cannot  be realized
within the Standard Model (SM)~\cite{EWSM}, since the phase transition
is  not sufficiently  strong enough  given  the direct  limits on  the
Higgs-boson  mass.  In  this case,  one  needs to  resort to  extended
models with extended Higgs sectors, such as the MSSM~\cite{EWMSSM}.

\item {\it Baryogenesis through leptogenesis}.  This scenario combines
the two key  ideas of the above two  scenarios.  Leptogenesis is based
on  the  out-of-equilibrium  $L$-violating  decays of  heavy  Majorana
neutrinos.   These decays  produce a  net lepton  asymmetry,  which is
converted  into the observed  BAU through  $(B+L)$-violating sphaleron
interactions~\cite{FY}.

\end{itemize}

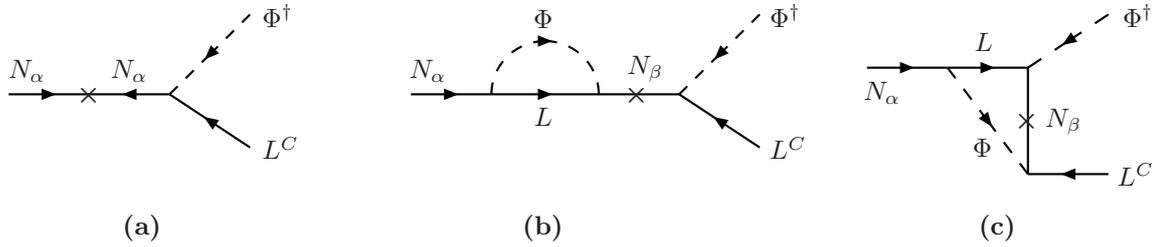
\begin{figure}[t]
{\small 
\begin{center}
\begin{picture}(450,100)(0,0)
\SetWidth{0.8}

\ArrowLine(20,60)(50,60)\ArrowLine(80,60)(50,60)
\Text(50,60)[]{{\boldmath $\times$}}
\ArrowLine(110,40)(80,60)\DashArrowLine(110,90)(80,60){5}
\Text(20,65)[bl]{$N_\alpha$}\Text(72,65)[br]{$N_\alpha$}
\Text(115,40)[l]{$L^C$}\Text(115,90)[l]{$\Phi^\dagger$}

\Text(70,10)[]{\bf (a)}

\ArrowLine(170,60)(200,60)\ArrowLine(200,60)(240,60)
\Line(240,60)(270,60)\Text(255,60)[]{{\boldmath $\times$}}
\DashArrowArcn(220,60)(20,180,0){5}
\ArrowLine(300,40)(270,60)\DashArrowLine(300,90)(270,60){5}
\Text(170,65)[bl]{$N_\alpha$}\Text(265,65)[br]{$N_\beta$}
\Text(220,85)[b]{$\Phi$}\Text(220,55)[t]{$L$}
\Text(305,40)[l]{$L^C$}\Text(305,90)[l]{$\Phi^\dagger$}

\Text(220,10)[]{\bf (b)}

\ArrowLine(340,70)(370,70)\ArrowLine(370,70)(400,70)
\Line(400,70)(400,30)\ArrowLine(430,30)(400,30)
\Text(401,50)[]{{\boldmath $\times$}}  
\DashArrowLine(430,90)(400,70){5}\DashArrowLine(370,70)(400,30){5}
\Text(340,65)[lt]{$N_\alpha$}\Text(385,80)[]{$L$}\Text(408,50)[l]{$N_\beta$}
\Text(437,90)[l]{$\Phi^\dagger$}\Text(435,30)[l]{$L^C$}
\Text(388,40)[r]{$\Phi$}

\Text(390,10)[]{\bf (c)}

\end{picture}
\end{center} }
\caption{Feynman diagrams contributing  to the $L$-violating decays of
heavy Majorana  neutrinos, $N_\alpha \to L^C  \Phi^\dagger$, where $L$
and   $\Phi$   represent    lepton   and   Higgs-boson   iso-doublets,
respectively: (a) tree-level graph, and (b) self-energy and (c) vertex
graph.}\label{f1}
\end{figure}

Several variants  of leptogenesis  have been discussed  so far  in the
literature~\cite{RNM}.  These  include Dirac leptogenesis~\cite{DLRW},
non-thermal     leptogenesis~\cite{Laz/Shafi},     Affleck-Dine    and
spontaneous leptogenesis~\cite{DK}, and RL that will be discussed more
extensively in Section~\ref{sec:RL}.

\section{The Flavourdynamics of Leptogenesis \label{sec:flavour}} 
\medskip

An important  recent development is  the understanding of  the special
role  that flavour  effects can  play in  leptogenesis. In  fact, this
special role  of flavour effects  on scenarios of baryogenesis  is not
new~\cite{KS}    and   has    already   been    recognised    in   the
past~\cite{Dreiner/Ross,CKO}.   The  key   observation  here  is  that
sphalerons do not only conserve $B-L$, but also the individual quantum
numbers:
$$\frac{1}{3}  B  -  L_{e,\mu,\tau}\;  .$$ Thus,  a  generated  baryon
asymmetry can be protected from erasure if a particular lepton flavour
combination or an individual  lepton number $L_{e,\mu,\tau}$ is out of
thermal  equilibrium.    For  instance,   this  idea  has   been  used
in~\cite{Dreiner/Ross}   to  show   how   specific  R-parity-violating
supersymmetric  scenarios do not  wash out  any baryon  asymmetry that
could eventually account for the observed matter--antimatter asymmetry
in the Universe.

\begin{figure}[t]
  \begin{center}
      \includegraphics[width=30pc]{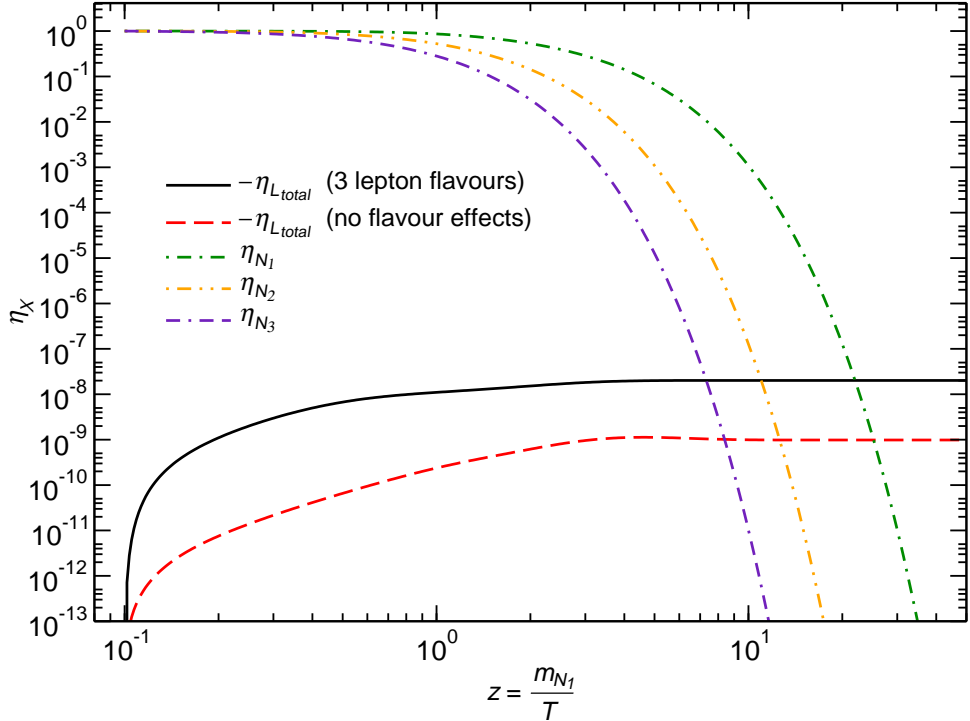}
  \end{center}
\caption{\label{LeptoGen} Numerical estimates for a scenario
with $m_{N_1} = 10^{10}~{\rm GeV}$, $m_{N_2} = 2m_{N_1}$, 
$m_{N_3} = 3 m_{N_1}$.}
\end{figure}

The new aspect of flavourdynamics  in leptogenesis is that the BAU can
now be both {\em generated from} and {\em protected in} a {\it single}
lepton   flavour~\cite{APtau},   e.g.~in   the  $\tau$-lepton   number
$L_\tau$. In order to better understand this, we need first to clarify
the two sources of flavour effects:
\begin{itemize}

\item  {\it flavour  effects  due to  charged-lepton Yukawa  couplings
$h_{e,\mu,\tau}$.}   These  effects are  related  to the  interactions
mediated  by charged-lepton  Yukawa  couplings~\cite{BCST}.  If  these
interactions are  in or  out of thermal  equilibrium at the  scale, at
which leptogenesis takes  place, then the predicted value  for the BAU
gets   modified  significantly,   typically   up  to   one  order   of
magnitude~\cite{NRR}. Thus,  if the leptogenesis scale  is higher than
$10^9$~TeV,  then all  charged-lepton Yukawa  interactions are  out of
thermal  equilibrium,  whereas  if  the  scale is  below  10~TeV,  all
charged-lepton Yukawa couplings  are in thermal equilibrium~\cite{CKO}
and   their   full  impact   on   the   BAU   needs  be   taken   into
account~\cite{PU2}.

\item   {\it   flavour    effects   due   to   heavy-neutrino   Yukawa
couplings~$h^\nu_{l\alpha}$.}    These   effects   are   governed   by
heavy-neutrino Yukawa-coupling  interactions~\cite{APtau,EMX}. In this
case, one may  consider scenarios that would generate  an excess in an
individual  lepton number,  e.g.~$L_\tau$, and  protect  the resulting
asymmetry from  potentially large wash-out effects  due to sphalerons.
For  example,   one  could  arrange  the   neutrino  Yukawa  couplings
$h^\nu_{l\alpha}$, such that  all heavy Majorana neutrinos $N_{1,2,3}$
decay  out   of  equilibrium  in  $L_{\tau}$,  $N_1$   decays  out  of
equilibrium  in  $L_{e,\mu,\tau}$,  but  $N_{2,3}$  could  decay  very
rapidly into $L_{e,\mu}$.  In Fig.~\ref{LeptoGen}, we give predictions
for an hierarchical scenario with and without neutrino Yukawa coupling
effects included~\cite{PU2,EMX}.  The two predictions can differ by up
to one order of magnitude. As we will discuss in Section~\ref{sec:RL},
however, flavour  effects due to~$h^\nu_{l\alpha}$ may  modify the BAU
predictions by  many orders of magnitude, e.g.~bigger  than $10^6$, in
RL      models,       thereby      rendering      their      inclusion
indispensable~\cite{APtau,PU2}.

\end{itemize}

We close this section by  mentioning that flavour effects do also play
an important role  in collision terms of the  Boltzmann equations that
describe scatterings~\cite{PU,PU2}. These are $\Delta L =1$ Yukawa and
gauge-mediated  scatterings, as  well as  $\Delta L  =0,2$ scatterings
mediated by heavy Majorana neutrinos.

\section{Resonant Leptogenesis \label{sec:RL}} 
\medskip

A   scenario  inspired   by   the  dynamics   of  the   $K^0\bar{K}^0$
system~\cite{LOY}  is resonant  leptogenesis~\cite{APRD,PU}.  Resonant
leptogenesis is  based on the observation  that self-energies dominate
the    lepton    asymmetries    if    $|m_{N_1}   -    m_{N_2}|    \ll
m_{N_{1,2}}$~\cite{LiuSegre}.   However,  as we  will  see below,  the
proper inclusion of heavy-neutrino width effects is necessary in order
to  obtain a  non-singular  well-behaved analytic  expression for  the
leptonic asymmetries.

The are several variants of RL:
\begin{itemize}

\item  {\it  Soft  RL~\cite{soft}.}   In this  scenario,  leptogenesis
results  from  sneutrino  decays   in  the  MSSM.   CP-violating  soft
SUSY-breaking  parameters may  split  the degeneracy  within a  single
generation of right-handed sneutrino states, which leads to a resonant
enhancement of the leptonic asymmetries.

\item {\it Radiative RL~\cite{rad}.}  In this case, the heavy Majorana
neutrinos are  exactly degenerate at  the GUT scale. Then,  small mass
differences among the heavy Majorana neutrino states are generated via
renormalization  group (RG)  running from  the GUT  scale down  to the
leptogenesis scale. These small mass differences give rise to RL.

\item     {\it      Coherent     RL     (via      sterile     neutrino
oscillations)~\cite{ARS,Misha}.}    This   scenario   relies  on   the
possibility that sterile neutrinos with masses well below the critical
temperature of the electroweak phase transition, e.g.~of order 10~GeV,
maintain  the coherence of  their CP  asymmetric oscillations  and can
thus produce  an enhanced leptonic  asymmetry of the amount  needed to
create the observed BAU.

\end{itemize}

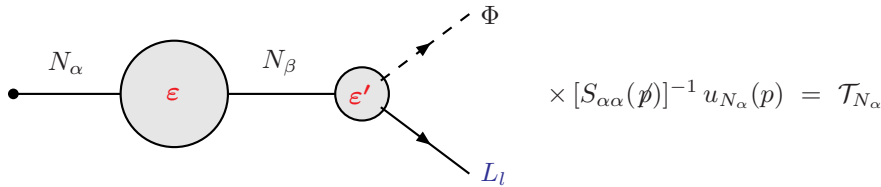
\begin{figure}[t]
{\small 
\begin{picture}(400,100)(0,0)
\SetWidth{0.8}

\Vertex(50,50){2}
\Line(50,50)(90,50)\Text(70,62)[]{$N_\alpha$}
\Line(130,50)(170,50)\Text(150,62)[]{$N_\beta$}
\GCirc(110,50){20}{0.9}\Text(110,50)[]{{\boldmath ${\color{Red}\varepsilon}$}}
\GCirc(180,50){10}{0.9}\Text(180,50)[]{{\boldmath ${\color{Red}\varepsilon'}$}}
\DashArrowLine(187,55)(220,80){5}\Text(225,80)[l]{$\Phi$}
\ArrowLine(187,45)(220,20)\Text(225,20)[l]{\color{Blue}$L_l$}
\Text(250,50)[l]{$\times\, [S_{\alpha\alpha} (\not\! p)]^{-1}\, 
u_{N_\alpha} (p)\ =\
{\cal T}_{N_\alpha}$}
\end{picture} }
\caption{LSZ-type approach to mixing and decay of heavy Majorana
  neutrinos~\label{fig:LSZ}}
\end{figure}

We   will   now   briefly    review   the   field   theory   developed
in~\cite{APRD,PU}  for thermal  RL.   This is  based  on an  effective
LSZ-type formalism~\cite{APNPB}  with the aim to  incorporate both the
mixing  and the  decay  of  the heavy  Majorana  neutrinos.  As  shown
in~\ref{fig:LSZ},  one  needs   to  resum  self-energy  graphs,  whose
absorptive  parts  regularize  the  tree-level mass  singularity  when
$m_{N_\alpha} \to  m_{N_\beta}$.  For a  two-generation heavy-neutrino
model, the resummed propagator matrix reads:
\begin{equation}
S_{\alpha\beta} (\not\! p)\ =\ \left(\! \begin{array}{cc}
\not\!p - m_{N_1} + \Sigma_{11}(\not\! p) & \Sigma_{12} (\not\!p) \\
\Sigma_{21}(\not\!p) & \not\!p - m_{N_2} + \Sigma_{22}(\not\! p)
\end{array}\!\right)^{-1}\; . 
\end{equation}
Extensive technical details of this formalism may be found in~\cite{APRD,PU}.

Making use of the  corresponding $K^0\bar{K}^0$ terminology, there are
two types of CP violation in the leptonic decays of the heavy
Majorana neutrinos: 
\begin{itemize}

\item {\it  $\varepsilon'$-type  CP   violation.}  In  this  case,  the
required CP  violation arises from the interference  of the tree-level
graph with the absorptive of the self-energy effects. 
\begin{equation}
  \label{epsprime}
\varepsilon'_{N_\alpha}\ =\ 
\frac{{\rm Im}\, (h^{\nu\dagger}\,h^\nu)^2_{\alpha\beta}}{
(h^{\nu\dagger}\,h^\nu)_{\alpha\alpha}\,(h^{\nu\dagger}\,h^\nu)_{\beta\beta} }\
\left(\frac{\Gamma_{N_\beta}}{m_{N_\beta}}\right)\
f\left(\frac{m^2_{N_\beta}}{m^2_{N_\alpha}}\right)\,,
\end{equation}
where 
$$\Gamma_{N_\beta}\                                                  =\
\frac{(h^{\nu\dagger}\,h^\nu)_{\beta\beta}}{8\pi}\   m_{N_\beta}$$  is
the   tree-level    decay   width   of   $N_\beta$,    and   $f(x)   =
\sqrt{x}[1-(1+x)\ln(1+1/x)]$    is    the   Fukugita--Yanagida    loop
function~\cite{FY}.

\item {\it $\varepsilon$-type CP violation.}
\begin{equation}
  \label{epsilon}
\varepsilon_{N_\alpha}\ =\ 
\frac{{\rm Im}\,(h^{\nu\dagger}\,h^\nu)^2_{\alpha\beta}}{
(h^{\nu\dagger}\,h^\nu)_{\alpha\alpha}\,
(h^{\nu\dagger}\,h^\nu)_{\beta\beta} }\ 
\left(\frac{\Gamma_{N_\beta}}{m_{N_\beta}}\right)\
\frac{ (m^2_{N_\alpha} - m^2_{N_\beta})\, m_{N_\alpha}\,
  m_{N_\beta} }{ 
(m^2_{N_\alpha} - m^2_{N_\beta})^2\, +\,
  m^2_{N_\alpha}\Gamma^{2}_{N_\beta}} 
\end{equation}
Note     that    $\varepsilon_{N_\alpha}$     dominate     over    the
$\varepsilon'_{N_\alpha}$     when    $m_{N_\alpha}-m_{N_\beta}    \ll
m_{N_{\alpha,\beta}}$.  In the  limit $m_{N_\alpha}  \to m_{N_\beta}$,
the  would-be singularity  is regularized  by the  decay width  of the
heavy neutrino $N_\beta$.

\end{itemize}

\begin{figure}[t]
  \begin{center}
   \includegraphics[width=30pc]{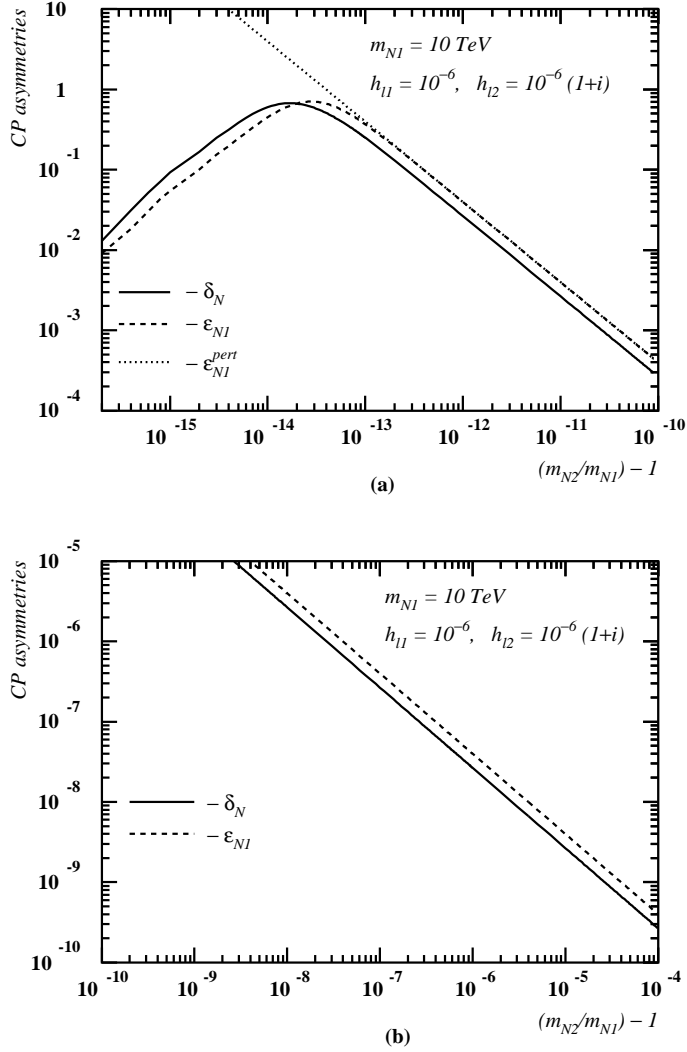}
\end{center} 
\vspace{-1cm}
\caption{$\varepsilon$- and $\varepsilon'$-type contributions to the
leptonic asymmetries in a model with two heavy Majorana neutrinos,
with $\delta_{N_{1,2}} \approx \varepsilon_{N_{1,2}} +
\varepsilon'_{N_{1,2}}$ and $\delta_N \approx \delta_{N_1} +
\delta_{N_2}$.}\label{fig:epsilon}
\end{figure}

From~(\ref{epsilon}),  the following {\it  two} conditions  for ${\cal
O}(1)$ leptonic asymmetries may be obtained~\cite{APRD}:
\begin{equation}
\mbox{(i)} ~~~m_{N_\alpha} - m_{N_\beta}\ \sim\ \frac{1}{2}\
\Gamma_{N_{\alpha,\beta}}\; ,\qquad 
\mbox{(ii)} ~~~\frac{\displaystyle {\rm Im}\,
(h^{\nu\dagger}\,h^\nu)^2_{\alpha\beta}}{\displaystyle
(h^{\nu\dagger}\,h^\nu)_{\alpha\alpha}\,(h^{\nu\dagger}\,h^\nu)_{\beta\beta}
}\ \sim\ 1\; .
\end{equation}
Figure~\ref{fig:epsilon} shows the impact  of the above two conditions
on the leptonic asymmetries for a low-scale seesaw model.  The generic
feature of the resonant conditions  remains valid even in the presence
of flavour effects~\cite{PU2}.

As was already mentioned in Section~\ref{sec:flavour}, flavour effects
due to neutrino  Yukawa couplings play an important  role in models of
RL. As  an illuminating  example, we will  consider a scenario  of RL,
which uses  the $\tau$-lepton number to  generate the BAU  and has the
potential to  be directly tested as  we will see in  the next section.
In such  a model, the smallness of  the light neutrinos is  not due to
the   usual  seesaw   mechanism~\cite{seesaw},  but   thanks   to  the
approximate         breaking         of         lepton         flavour
symmetries~\cite{MV,BGL,APZPC,APtau}.   In detail, the  heavy neutrino
sector is initially  assumed to be SO(3), and gets  broken down to the
identity   {\bf  I}  via   the  neutrino   Yukawa  couplings   in  two
steps~\cite{APtau}:
\begin{center}
SO(3) ~~$\stackrel{\sim\,
  h_\tau}{\longrightarrow}$~~ SO(2) $\simeq$ U(1)$_l$
~~$\stackrel{\sim\, h_e}{\longrightarrow}$~~ {\bf I}\; .
\end{center}
The neutrino  Dirac mass  matrix that results  from the  Yukawa sector
reads:
\begin{equation}
m_D\ =\ \frac{v}{\sqrt{2}}\,\left(\! \begin{array}{ccc}
\varepsilon_e & a\, e^{-i\pi/4} & a\, e^{i\pi/4} \\ 
\varepsilon_\mu & b\, e^{-i\pi/4} & b\, e^{i\pi/4} \\
\varepsilon_\tau & c\, e^{-i\pi/4} &c\, e^{i\pi/4} \end{array} \!\right)\, ,
\end{equation} 
with $a  \sim b  \sim 10^{-2} \sim  h_\tau$ and $c,\  |\varepsilon_l |
\sim 10^{-7} \sim h_e$. Likewise, the heavy Majorana mass matrix $m_M$
deviates from ${\bf  I}$ by terms $\Delta m_M \sim  ( h^{\nu T} h^{\nu
*}  +  h^{\nu \dagger}  h^\nu  )\, m_N$,  as  naively  expected by  RG
effects, where $m_N$ is an  SO(3) symmetric Majorana mass in the range
0.1--1~TeV.  Specifically, the light neutrino mass matrix is given
by~\cite{PU2} 
\begin{displaymath}
m^{\rm {\color{Black}light}}_\nu\ =\ 
\frac{v^2}{2m_N}\,\left(\! \begin{array}{ccc}
\frac{{\color{Black}\Delta m_N}}{m_N}\,a^2 - {\color{Black}\varepsilon^2_e}  & 
\frac{{\color{Black}\Delta m_N}}{m_N}\,ab - 
{\color{Black}\varepsilon_e\varepsilon_\mu} & 
\frac{{\color{Black}\Delta m_N}}{m_N}\,ac -
{\color{Black}\varepsilon_e\varepsilon_\tau} \\
\frac{{\color{Black}\Delta m_N}}{m_N}\,ab -
{\color{Black}\varepsilon_e\varepsilon_\mu}  & 
\frac{{\color{Black}\Delta m_N}}{m_N}\,b^2-{\color{Black}\varepsilon^2_\mu}  & 
\frac{{\color{Black}\Delta m_N}}{m_N}\,bc-
{\color{Black}\varepsilon_\mu\varepsilon_\tau}  \\
\frac{{\color{Black}\Delta m_N}}{m_N}\,ac -
{\color{Black}\varepsilon_e\varepsilon_\tau}  & 
\frac{{\color{Black}\Delta m_N}}{m_N}\,bc - 
{\color{Black}\varepsilon_\mu\varepsilon_\tau} & 
\frac{{\color{Black}\Delta m_N}}{m_N}\,c^2 - 
{\color{Black}\varepsilon^2_\tau} \end{array} \!\right)\; ,
\end{displaymath}
where 
$${\color{Black}\Delta m_N}\ =\ 2 ({\color{Black}\Delta m_M})_{23} + 
i [({\color{Black}\Delta m_M})_{33} -
({\color{Black}\Delta m_M})_{22}],\quad \frac{b}{a}\ =\ \frac{19}{50}\ ,$$
and (in $\sim 10^{-7}$ units)
$$
\sqrt{\frac{{\color{Black}\Delta m_N}}{m_N}}\,a \ =\  2\,,\quad 
{\color{Black}\varepsilon_e} \ =\ 2\: +\: \frac{21}{250}\; ,\quad
{\color{Black}\varepsilon_\mu} \ =\ \frac{13}{50} \; ,\quad 
{\color{Black}\varepsilon_\tau} \ =\ -\, \frac{49}{128}\; .
$$  
The above choice  of parameters  predicts a  light-neutrino sector
that  realizes an inverted  mass hierarchy,  $m_{\nu_3} <  m_{\nu_1} <
m_{\nu_2}$, with
\begin{eqnarray}
m_{\nu_2}^2 - m_{\nu_1}^2 \!& = &\! 7.54 \times
10^{-5}\,\,\mathrm{eV}^2\,,\qquad m_{\nu_1}^2 - m_{\nu_3}^2 \ =\ 
2.45 \times 10^{-3}\,\,\mathrm{eV}^2\,,\nonumber\\ 
\sin^2 \theta_{12} \! & = &\! 0.362 \,,\qquad
\sin^2 \theta_{23} \ = \ 0.341 \,,\qquad
\sin^2 \theta_{13} \ = \ 0.047 \;.\nonumber
\end{eqnarray}
The above  is compatible with low-energy neutrino  oscillation data at
the 3$\sigma$ level~\cite{JVdata}.

\begin{figure}[t]
  \begin{center}
      \includegraphics[width=30pc]{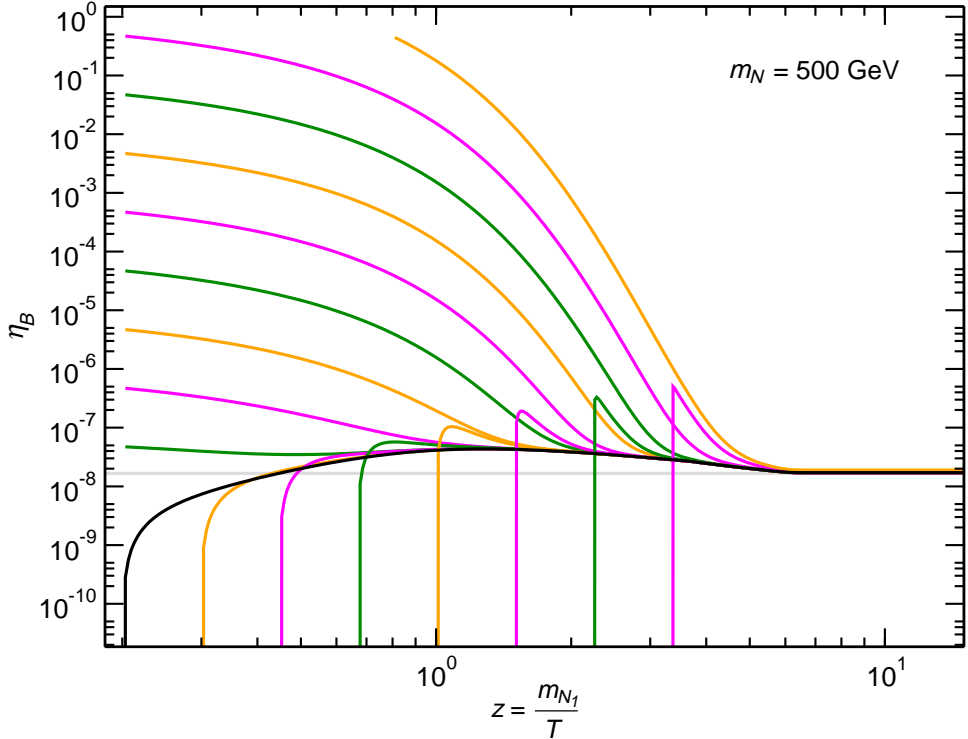}
  \end{center}
\caption{Numerical estimates of the baryon asymmetry $\eta_B$ for
  different initial conditions in the R$\tau$L model with
  $m_N=500$~GeV.  The horizontal line indicates the observed baryon
  asymmetry as required just after the sphalerons
  freeze out.\label{fig:500GeV}}
\end{figure}

\begin{figure}[t]
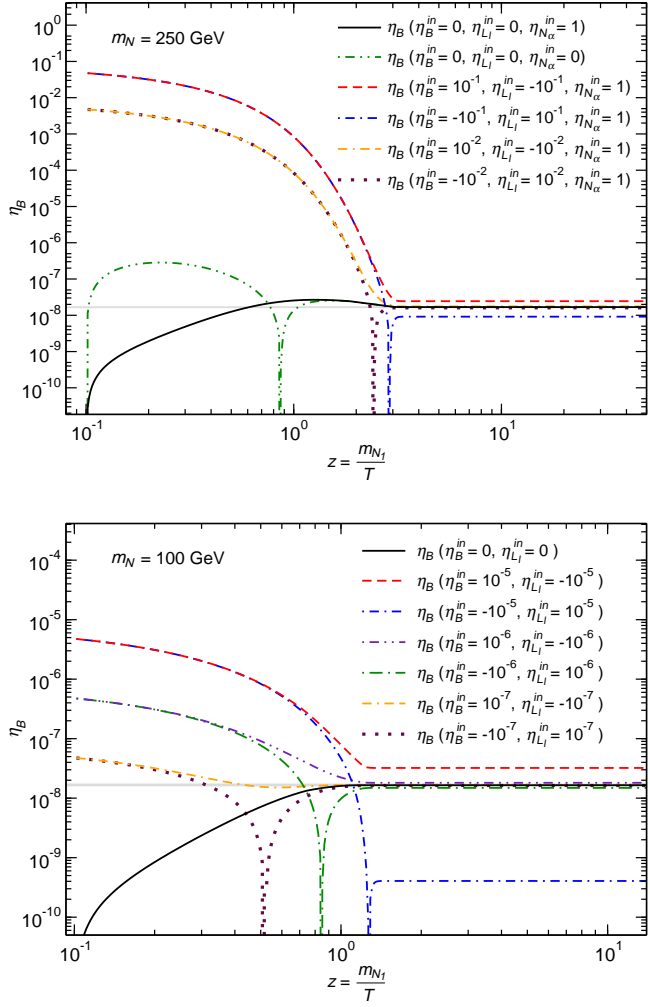

  \begin{center}
      \includegraphics[width=20pc]{250GeV-initial.eps}\\[0.5cm]
      \includegraphics[width=20pc]{100GeV-initial.eps}
  \end{center}
\caption{$\eta_B$ versus $z=m_{N_1}/T$ in the R$\tau$L model with
  $m_N=250$~GeV~(upper panel) and $m_N =100$~GeV~(lower panel).}
\label{fig:250GeV}
\end{figure}

In Fig.~\ref{fig:500GeV},  we give  numerical estimates of  the baryon
asymmetry $\eta_B$  for different  initial conditions in  the resonant
$\tau$-leptogenesis  (R$\tau$L)  model  described  above.   The  heavy
Majorana  mass scale  is  $m_N=500$~GeV.  Our  results  show that  the
baryon  asymmetry  $\eta_B$  is  almost  independent  on  the  initial
conditions. This property is a consequence of the fixed-point dynamics
which exhibits a system that  is in quasi-thermal equilibrium.  As can
be  seen from the  upper panel  in Fig.~\ref{fig:250GeV},  this almost
independence of $\eta_B$ on  the initial conditions persists, even if
the heavy Majorana mass $m_N$ is  as low as 250~GeV. For lower masses,
e.g.~$m_N = 100$~GeV (see  lower panel in Fig.~\ref{fig:250GeV}), this
feature becomes  less pronounced, and the dynamics  of leptogenesis at
the electroweak phase transition  becomes more involved. In this case,
one  needs to take  into consideration  contributions to  the leptonic
asymmetries that  originate from  the transverse polarizations  of the
$W$- and $Z$-boson gauge fields~\cite{APRD2}.

It is therefore tempting to introduce some measure that could quantify
the initial condition dependence  of the different models suggested in
the literature. To this end, we may define a $Q$-factor
\begin{equation}
  \label{Qfactor}
Q\ =\ 
\ln \left|\frac{\delta \eta_B^{\rm
 in}}{\delta \eta_B^{\rm fin}}\right|\; ,
\end{equation}
where  $T_{\rm  in}$  is   a  typical  initial  temperature  when  the
baryogenesis mechanism becomes active. For instance, for leptogenesis,
this could be  identified as $T_{\rm in} \sim  m_N$ or for electroweak
baryogenesis,  one  may  assume  that  $T_{\rm in}$  is  the  critical
temperature    of   the    electroweak   phase    transition   $T^{\rm
EW}_c$. Correspondingly,  $T_{\rm fin}$ is  the freeze-out temperature
of the baryogenesis mechanism under study. In terms of the $Q$-factor,
one could quantify the actual dependence of the model for baryogenesis
on the initial  conditions. Thus, if $Q \gg 1$,  this would imply that
there is very weak dependence of  the BAU on  the initial conditions,
whereas for  $Q <  0$ this  would signify that  the BAU  does strongly
depend on these. Evidently, RL models, as the one analyzed here, have 
large values of $Q$ and so render them quite predictive.

\section{Particle-Physics Phenomenology of Resonant Leptogenesis 
     \label{sec:PPP}}
\medskip

RL models  can give  rise to a  number of  phenomenologically testable
signatures.  Here, we will present the generic predictions of R$\tau$L
models  for the $0\nu\beta\beta$  decay, and  for processes  of lepton
flavour violation (LFV), such as  $\mu \to e\gamma$, $\mu \to eee$ and
$\mu  \to  e$  conversion  in  nuclei.   Finally,  we  give  numerical
estimates of production cross  sections of heavy Majorana neutrinos at
the LHC.

\subsection{{\boldmath $0\nu\beta\beta$} Decay}
\medskip

Neutrinoless  double beta  decay  ($0\nu\beta\beta$) corresponds  to a
process in which  two single  $\beta$ decays~\cite{doi85}
occur simultaneously  in  one nucleus.  As  a consequence  of this,  a
nucleus $(Z,A)$ gets converted into a nucleus $(Z+2,A)$, i.e.\
\begin{displaymath}        
^{A}_{Z}\,X\ \to\ ^A_{Z+2}\,X\: +\: 2 e^-\; . 
\end{displaymath}
Evidently,  this process  violates  $L$-number by  two  units and  can
naturally take place in minimal RL models, in which the observed light
neutrinos are  Majorana particles.  The observation of  such a process
would  provide  further information  on  the  structure  of the  light
neutrino mass matrix ${\bf m}^\nu$. In particular, the transition rate
of  $0\nu\beta\beta$  decay  is  governed by  the  effective  Majorana
neutrino  mass  $\langle  m  \rangle$.  For the  R$\tau$L  model,  the
effective neutrino mass is given by
\begin{equation}
|\langle m \rangle|\ =\  |({\bf m}^\nu)_{ee}|\ =\ 
\frac{v^2}{2m_N}\: \bigg|\,\frac{\Delta m_N}{m_N}\,a^2\ -\
\varepsilon^2_e\,\bigg|\ \approx\ 0.013~{\rm eV}\; .\; . 
\end{equation}
Such a  prediction lies within  the reach of future  experiments which
will  be  sensitive  to  values  of $|\langle  m  \rangle|$  of  order
$10^{-2}$, such as SuperNEMO~\cite{SuperNEMO}.

\subsection{$\mu \to e\gamma$, $\mu \to eee$ and $\mu \to e$ conversion}
\medskip

Quantum  effects  due  to  heavy  Majorana neutrinos  may  induce  LFV
couplings to the photon ($\gamma$) and the $Z$ boson.  These couplings
give rise to LFV decays, such as $\mu \to e\gamma$~\cite{CL}, $\mu \to
eee$~\cite{IP} and $\mu \to e$ conversion~\cite{BNT}.  More details on
the   actual   calculation   of   these   processes   may   be   found
in~\cite{IP,Ara}.   Also,  related  phenomenological analyses  of  LFV
effects   in   the   SM   with   singlet  neutrinos   may   be   found
in~\cite{LFVrev,KPS,BKPS,Ofit}.

Let us  first consider the decay  $\mu \to e\gamma$.   In the R$\tau$L
model, the branching fraction for $\mu \to e\gamma$ is given by
\begin{equation}
  \label{Bllgamma}
B(\mu \to e \gamma )\ =\ 9 \cdot 10^{-4}\, \times\,
\frac{|a|^2\,|b|^2\, v^4}{m^4_N}\ .
\end{equation}
The theoretical prediction should  be contrasted with the experimental
upper limit:
\begin{equation}
  \label{Bexpllgamma}
B_{\rm exp} (\mu \to e \gamma)\ <\ 1.2\, \times 10^{-11}\, .\quad
\end{equation}
For  the particular  scenario considered  in  Section~\ref{sec:RL}, we
find  $B(\mu \to  e \gamma  ) \sim  10^{-12}$. These  values  are well
within reach of  the MEG collaboration, which will  be sensitive to $B
(\mu \to e\gamma ) \sim 10^{-14}$~\cite{MEG}.

We then consider the decay  $\mu \to eee$.  Its branching ratio $B(\mu
\to eee )$ may be related to $B(\mu \to e\gamma)$ through~\cite{PU2}:
\begin{equation}
  \label{Rmueee}
B (\mu\to e e e)\ \simeq\  
8.2 \cdot 10^{-3}\, \times
\Bigg[\,1\ -\ 0.8\ln\bigg(\frac{m^2_N}{M^2_W}\bigg)
\ +\ 0.5 \ln^2\bigg(\frac{m^2_N}{M^2_W}\bigg)
\, \Bigg]\: B(\mu\to e\gamma)\; .
\end{equation}
For  example,    for   an R$\tau$L   model    with  $m_N  =  250$~GeV,
(\ref{Rmueee}) implies
\begin{equation}
B (\mu\to e e e)\ \simeq\ 1.4\cdot 10^{-2}\,\times B(\mu \to e\gamma)\
\simeq\ 1.4\cdot 10^{-14}
\end{equation}
This  value is  a  factor  $\sim 70$  below  the present  experimental
bound~\cite{PDG}: $B_{\rm  exp} (\mu \to eee) <  1.0 \times 10^{-12}$.
It would  be therefore encouraging, if  higher sensitivity experiments
could be designed to further probe this observable.

Finally, one  of the  most sensitive experiments  to probe LFV  is the
coherent conversion  of $\mu \to  e$ in nuclei. Specifically,  for the
${}^{48}_{22}{\rm Ti}$ case, $B_{\mu  e} (26,22)$ can be also directly
linked to $B(\mu \to e\gamma)$ through the relation~\cite{PU2}:
\begin{equation}
  \label{Rmue}
B_{\mu e} (26,22)\ \simeq\  
0.1\,\times \Bigg[\,1\ +\
0.5\ \ln\bigg(\frac{m^2_N}{M^2_W}\bigg)\Bigg]^2\:
B(\mu\to e\gamma)\; .
\end{equation}

On  the  experimental  side,  the  upper  bound on  the  $\mu  \to  e$
conversion rate in ${}^{48}_{22}{\rm Ti}$~\cite{SINDRUM} is
\begin{equation}
  \label{mueconvexp}
B^{\rm exp}_{\mu  e} (26,22)\ <\ 4.3\,\times 10^{-12}\, , 
\end{equation}
at    the    90\%    CL.     However,    the    proposed    experiment
PRISM/PRIME~\cite{PRISM}  will  be sensitive  to  conversion rates  of
order $10^{-18}$.

Given~(\ref{Rmue}) and $m_N = 250$~GeV, the prediction for $\mu \to e$
conversion in ${}^{48}_{22}{\rm Ti}$ is:
\begin{equation}
B_{\mu e} (26,22)\  \simeq\    0.46 \times B(\mu \to   e\gamma)\ \sim\
4.5 \times 10^{-13}\;.
\end{equation}
The  above  prediction falls well    within reach of  the  sensitivity
proposed by the PRISM/PRIME collaborations.

As opposed to other  scenarios of baryogenesis, leptogenesis models in
general do not suffer from  predicting a too large contribution to the
electron electric  dipole moment (EDM).   This is because  EDM effects
first arise at two loops and are suppressed either by higher powers of
small Yukawa  couplings and/or by  small factors, such as  $(m_{N_1} -
m_{N_{2,3}})/m_N$   in  RL  models~\cite{APRD,NN}.    Finally,  other
manifestations  of  heavy  Majorana   neutrino  effects  are  LFV  and
universality-breaking effects  in the $Z$-boson decays~\cite{KPS,BKPS}
and non-unitarity effects in neutrino oscillations~\cite{Malinsky}.

\subsection{Heavy Majorana Neutrino Production at the LHC}
\medskip

\begin{figure}[t]
  \begin{center}
    \includegraphics[width= 30pc]{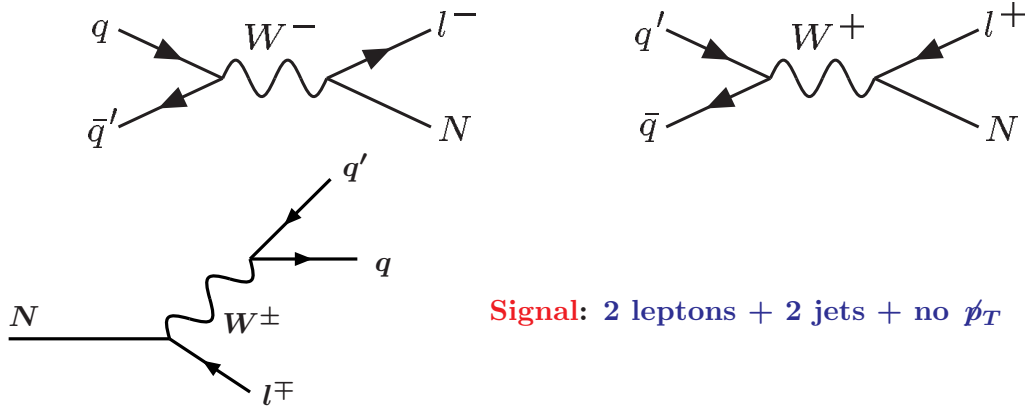}
  \end{center}
\begin{center}
\begin{picture}(450,120)(0,0)
\SetWidth{1.2}

\Line(20,60)(80,60)
\ArrowLine(110,40)(80,60)\Photon(110,90)(80,60){6}{2}
\ArrowLine(110,90)(150,90)
\ArrowLine(140,120)(110,90)
\Text(20,65)[bl]{\boldmath $N$}
\Text(115,40)[l]{\boldmath $l^\mp$}\Text(100,67)[l]{\boldmath $W^\pm$}
\Text(157,87)[l]{\boldmath $q$}\Text(145,125)[l]{\boldmath $q'$}

\Text(200,70)[l]{\bf {\color{Red}Signal}: {\color{Blue}2 leptons + 2 jets + no
  {\boldmath $\not\!p_T$}}}
\end{picture}
\end{center}
\vspace{-1cm}
\caption{Production and decay of heavy Majorana neutrinos at the LHC.}
\label{fig:LSD}
\end{figure}

Electroweak-scale RL models, such as the R$\tau$L model, predict heavy
Majorana  neutrinos with masses  and couplings  within the  range that
could  be produced  at the  LHC.  The  dominant production  channel is
through     the    $W^\pm$-boson     exchange    graph     shown    in
Fig.~\ref{fig:LSD}~\cite{APZPC,DGP,HZ,AAP,BEHS}.   The  characteristic
signature is  the observation of  like-sign dileptons with  no missing
transverse momentum~\cite{Wai/Goran}.

According to  the recent studies~\cite{HZ,AAP}, the LHC  may be rather
sensitive  to  the  light-to-heavy  neutrino  mixing  expression  $b^2
v^2/(2m^2_N)  \sim  10^{-3}$--$10^{-4}$,  for  $m_N  =  100$--150~GeV.
Hence, a  possible discovery  of heavy Majorana  neutrinos at  the LHC
will unravel  the scale of lepton-flavour violation,  thus putting the
predictions of  R$\tau$L models into direct  test.  Finally, low-scale
seesaw models of the type  considered here could also predict sizeable
signatures of simultaneous lepton flavour and lepton number violation,
such as  $pp \to  e^+\mu^+,\ e^-\mu^-,\ e^-\tau^-  \dots$, as  well as
resonantly enhanced ${\cal O}(1)$ CP  asymmetries~\cite{BLP}.  Thus, a
complete exploration of the phenomenological consequences of RL models
would be interesting.

\section{Conclusions  \label{sec:Concl}}
\medskip

Leptogenesis  provides  an interesting  mechanism  for explaining  our
observable matter--antimatter asymmetric Universe, which may relate to
the origin of  the neutrino mass.  In this brief  review, we have paid
particular attention to the recent  developments on the field, such as
the flavourdynamics of leptogenesis and resonant leptogenesis near the
electroweak phase transition.  In particular, flavourdynamics plays an
important  role   in  the   predictions  for  the   BAU  and   in  the
model-building of phenomenologically  testable scenarios.  It has been
explicitly demonstrated how  the observed matter--antimatter asymmetry
in  the Universe  may be  successfully explained  by  thermal resonant
leptogenesis at  the electroweak scale, independently  of any non-zero
primordial baryon asymmetry. In  fact, any primordial baryon asymmetry
will be  erased and the  actual value  of the BAU  will be set  by the
actual  mechanism itself,  as  a consequence  of the  quasi-in-thermal
equilibrium conditions that govern resonant leptogenesis.

Models of resonant leptogenesis  may have rich collider phenomenology.
Electroweak-mass heavy Majorana neutrinos  may give rise to observable
signatures  of  lepton number  and  lepton  flavour  violation at  the
LHC. There may also give rise to correlated predictions for low-energy
observables,  such  as $B(\mu  \to  e\gamma)$,  coherent  $\mu \to  e$
conversion in nuclei, $B(\mu \to eee)$ and the effective neutrino mass
in  the   $0\nu\beta\beta$  decay.  Thus,  an  exciting   era  of  new
discoveries at  the LHC  and in other  low-energy experiments  may lie
just ahead  of us, revealing  a greater picture that  unifies particle
physics and cosmology.

\section*{Acknowledgements}
I  am grateful  to the  organisers of  the conference  DISCRETE'08 for
their hospitality.

\end{document}